# High Speed and High Resolution Table-Top Nanoscale Imaging

G.K. Tadesse[1,2,*], R. Klas[1,2], S. Demmler[1,2], S. Hädrich[1,2], I. Wahyutama[1,2], M. Steinert[2], C. Spielmann[1,3], M. Zürch[1,3,4], A. Tünnermann[1,2,5], J. Limpert[1,2,5] and J. Rothhardt[1,2]

[1]*Helmholtz-Institute Jena, Fröbelstieg 3, 07743 Jena, Germany*
[2] *Institute of Applied Physics, Abbe Center of Photonics, Friedrich-Schiller-Universität Jena, Albert-Einstein-Straße 15, 07745 Jena, Germany*
[3]*Institute of Optics and Quantum Electronics, Abbe Center of Photonics, Friedrich-Schiller-Universität Jena, Max-Wien-Platz 1, 07743 Jena, Germany*
[4]*University of California, Department of Chemistry, CA 94720, Berkeley, USA*
[5]*Fraunhofer Institute for Applied Optics and Precision Engineering, Albert-Einstein-Str. 7, 07745 Jena, Germany*
*Corresponding author: getnet.tadesse@uni-jena.de*

**Abstract**
We present a table-top coherent diffraction imaging (CDI) experiment based on high-order harmonics generated at 18 nm by a high average power femtosecond fiber laser system. The high photon flux, narrow spectral bandwidth and high degree of spatial coherence allow for ultra-high sub-wavelength resolution imaging at a high numerical aperture. Our experiments demonstrate a half-pitch resolution of 13.6 nm, very close to the actual Abbe-limit of 12.4 nm, which is the highest resolution achieved from any table-top XUV or X-ray microscope. In addition, 20.5 nm resolution was achieved with only 3 sec of integration time bringing live diffraction imaging and 3D tomography on the nanoscale one step closer to reality. The current resolution is solely limited by the wavelength and the detector size. Thus, table-top nanoscopes with only a few-nm resolutions are in reach and will find applications in many areas of science and technology.

Coherent diffractive imaging (CDI) is an imaging technique that provides amplitude and phase information of a nanoscale sample from diffraction patterns recorded in the far field. Since no optics is needed between the sample and the detector, it is scalable to smallest resolutions provided that a high photon flux short wavelength light source with good coherence is used for illumination. Despite huge technological efforts, the resolution of conventional X-ray microscopes is still limited to 12 nm to 20 nm [1–3] by the fabrication precision of the employed zone plates. In contrast, coherent diffractive imaging and related techniques demonstrated 7 nm [4] and 5 nm resolution [5] already which can be improved with the availability of a better source. Since the short wavelength light can, in contrast to electron beams, even shine through μm-thick samples exciting possibilities in damage-free 3-dimensional (3D) imaging with unprecedented resolution open up [6]. Furthermore, ultrashort X-ray pulses enable time-resolved movies of the fastest dynamics on the nanoscale [7] being relevant for future electronic, optical and magnetic devices. Unfortunately, the applicability of these imaging techniques in science and technology is limited due to the size, cost and accessibility of the typically desired light sources namely synchrotrons and free-electron lasers [8].

The advantages of coherent nanoscale microscopy can only be fully exploited in all areas of science with a compact, reliable and powerful table-top implementation. Thus, laser-driven light sources based on high harmonic generation (HHG) [9,10] are considered as a promising alternative which can be implemented on a table-top in a standard research laboratory environment [6]. Previous table-top coherent imaging experiments have demonstrated 22 nm [11] and sub-wavelength [12] spatial resolution. Unfortunately, due to the limited photon flux of the table-top XUV sources measurement times of tens of minutes were required for imaging with such systems so far.

Clearly, real-world applications in nanoscience require shorter integration times and the smallest possible resolutions. Once the measurement times for a single high-resolution 2D image has been reduced to seconds, even 3D tomography or ptychographic imaging of large objects [13,14], which requires imaging of hundreds of individual diffraction patterns, get practically feasible with table-top setups. Significantly shorter integration times will also enable high through-put applications in life sciences, e.g. for morphology analysis and subsequent classification of cells [15]. Obviously, this requires a short illumination wavelength and a higher photon flux from the light source at the same time. HHG sources have generated coherent keV radiation already, but suffer from a dramatic decrease of the generated photon flux for high photon energies / short wavelengths [16,17]. Thanks to the recent advances in high average power fiber lasers, sources in the XUV and soft X-ray region based on HHG can nowadays provide orders of magnitude higher photon flux [18,19].

Here, we present a coherent diffractive imaging system employing a high photon flux 68.6 eV (18 nm) HHG source. The operating wavelength has been chosen as a compromise between high photon flux and short wavelength so as to enable sub-wavelength imaging with a high numerical aperture (NA=0.7). We achieved a spatial resolution of 13.6 nm, which represents a record for table-top imaging systems. In addition, 20.5 nm resolution has been achieved with only 3 s of integration time which enables real-time imaging and is a pre-requisite for table-top 3D nanoscale tomography and recording of nanoscale-movies of ultrafast processes.

A two-channel coherently combined fiber CPA (CC-FCPA) system similar to the one reported in [20] is used to deliver pulses having 320 fs duration with 1 mJ energy at a repetition rate of 30 kHz. The output pulses from the CC-FCPA at a central wavelength of 1030 nm are nonlinearly-compressed to 33 fs pulse duration with 0.55 mJ pulse energy [20]. The high-order harmonic generation (HHG) setup uses a 250 mm lens to focus the compressed pulses to a FWHM diameter of 50 μm in front of a gas jet delivering Argon at up to 4 bar of backing pressure into a vacuum chamber. This results in a peak intensity of about $3 \times 10^{14}$ W/cm$^2$ assuming a Gaussian shaped temporal and spatial profile for the driving pulsed beam. During the experiment, the intensity at the focus was adjusted for maximum XUV flux and good XUV beam profile using an aperture located just before the focusing lens which reduced the intensity by ~30% at the optimum opening. The generated XUV radiation was separated from the driving laser beam using a pair of grazing incidence plates, which reflect 40 % of the XUV but transmit most of the infrared light [21]. Two additional 200 nm thick Al filters suppress the remaining infrared light and the XUV spectrum is then measured using a grating spectrometer. Integrating over the 57$^{th}$ harmonic, a photon flux of about $1 \times 10^{10}$ photons /sec was generated at the gas jet around 68.6 eV.

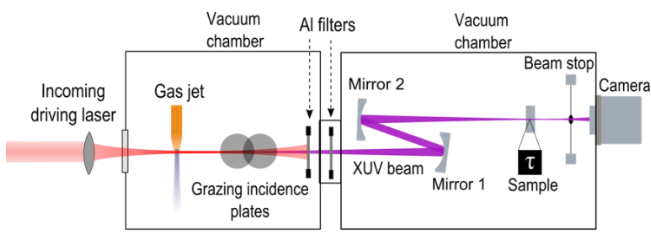

Fig. 1. Top view of the experimental setup of the HHG and imaging chambers. The grazing incidence plates are placed 11mm apart vertically with angle of incidence 82° to the surface normal.

The CDI setup is shown in Fig. 1 and uses two focusing mirrors (optiX fab) each having peak transmission of 50% at 68.6 eV to focus the XUV beam on the sample to be imaged. The energy bandwidth of the mirrors is narrow enough (2.2 eV FWHM) to suppress the neighboring harmonic lines by approximately one order of magnitude. The angle of incidence on these mirrors was set as small as possible (2.5° to the normal) to reduce astigmatism on the focused XUV beam. The XUV beam has a Gaussian-like far-field intensity profile at the camera that can be optimized by adjusting the driving laser intensity and gas jet position relative to the focus. The intensity FWHM of the XUV beam at the focus is estimated to be 10 μm by scanning a 1μm pinhole through it.

The sample is placed at the focal point of the XUV beam so as to have a flat phase front and a maximum photon flux available for imaging. The XUV camera (Andor iKon-L) has 2048 by 2048 pixels with each pixel being 13.5 μm wide. The distance between the sample and camera is 13.6 mm which allows for a high numerical aperture of 0.7. In order to evaluate the degree of coherence of the XUV beam, a diffraction pattern from a double slit sample with 1.5 μm separation was measured as shown in Fig. 2. The diffraction pattern shows an excellent contrast with fringe visibility of ~90 % which demonstrates the high degree of spatial coherence of the XUV beam.

The achievable resolution for a CDI setup is additionally limited by the temporal coherence of the source to [22]

$$\Delta r \geq O a \frac{\Delta\lambda}{\lambda} \qquad (1)$$

where $a$ is the sample size, and $O$ is the oversampling degree ($O = \sqrt{\sigma}$ for 2D imaging). To allow for a reliable reconstruction, the oversampling ratio ($\sigma$) must be larger than 2 and higher values are beneficial to increase the signal to noise ratio. The relative bandwidth, $\Delta\lambda/\lambda$, at the 57$^{th}$ harmonic (68.6 eV) has been measured to be ~1/200. Thus, the temporal coherence limit $\Delta r$ for our measurement is only 11.5 nm which is below the Abbe limit implying that our resolution is limited only by the NA of the setup, hence the detector size.

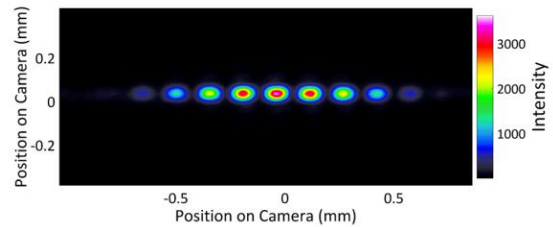

Fig. 2. Diffraction from a double slit sample with 0.1 sec. exposure time for a double slit with separation of 1.5 μm.

A transmission sample shown in Fig. 3(b) was fabricated by etching an aperture into a 50nm thick $Si_3N_4$ membrane by means of focused ion beam milling and subsequent coating with 150nm of Au by thermal evaporation. The sample size (1 μm) is smaller than the slit separation used in Fig. 2 and visibility > 90 % is to be expected between waves diffracted from opposite edges of the sample. Several diffraction patterns with acquisition times varying from 1 sec. to 15 min. were recorded and the bright central part was blocked by a beam stop for measurements with acquisition times longer than 10 sec. For each measurement, 8 by 8 pixels were binned together directly at the CCD which resulted in oversampling ratio of 5.2. These patterns were then merged together by multiplying each pattern with an appropriate scaling factors found from the usable overlapping regions of the individual measurements [12]. This increases the dynamic range of the measurement to more than seven orders of magnitude. In addition, curvature correction of the measured diffraction pattern was performed due to the high numerical aperture of the setup. Fig. 3(a) shows the final combined diffraction pattern after curvature correction. Diffraction speckles can be clearly seen until the very edge with some parts of the pattern even being cut at the edge of the detector.

For the phase retrieval, a guided version [23] of the RAAR algorithm [24] (β parameter 0.95) without any a priori knowledge of the sample except that it is an isolated object was employed. After 250 iterations the support was dynamically updated using the shrink-wrap method [25], while an initial support was generated by thresholding the autocorrelation of the measured diffraction pattern at 12% of its peak signal and subsequently using locator sets for refinement of the first support estimate [26]. We started with ten independent reconstructions and better results were achieved with limiting the real space to positive values. After the final run all reconstructions were averaged with the best of the reconstructions judged by the error metric [27]. This procedure was repeated until all reconstructions reached a stable solution. The amplitude of the reconstructed object (Fig. 3(c)) shows that every feature of the object is faithfully reconstructed. Differences from the SEM image (Fig. 3 (b)) can especially be seen in the central ~0.9 µm long and < 100 nm wide bar but also at other edges. This can be caused by non-perfect edges or waveguide effects [28] due to the small size (only a few wavelengths) and the high aspect ratio of the sample's features.

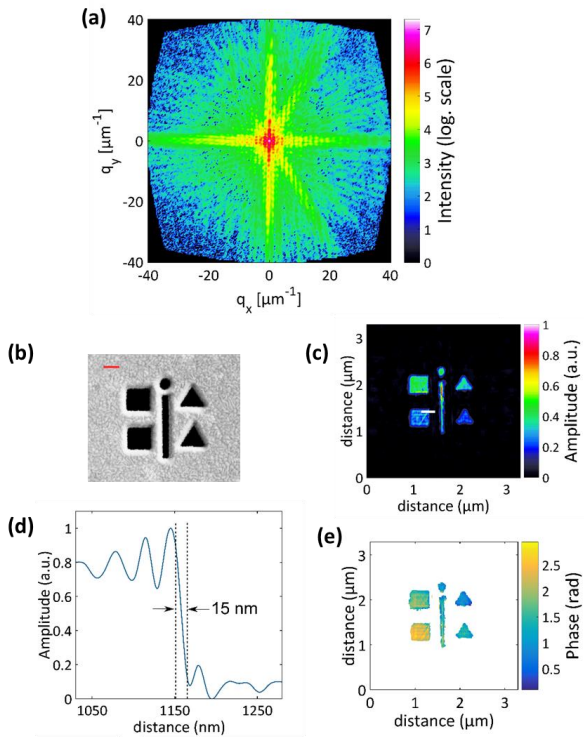

Fig. 3. (a) Curvature corrected high dynamic range diffraction pattern of the sample. (b) SEM picture of the transmission sample – scale bar is 200 nm wide. Reconstructed object (c) amplitude and (e) phase. (d) Cross-section taken along the white line in (c) shows 10/90 % resolution of 15nm.

A reliable and established measure of the highest spatial frequency that contributes to the reconstruction is the phase retrieval transfer function (PRTF) [29]. The half-pitch resolution, as defined by the PRTF 1/e criterion [30], is then the feature size that corresponds to the highest spatial frequency having PRTF value greater than 1/e. The PRTF analysis for our reconstructions is shown in Fig. 4 (blue line). It shows values greater than 1/e for almost all measured spatial frequencies. This is to be expected as the diffraction intensity was above the noise level for the whole camera with good contrast. Spatial frequencies up to 36.6 µm$^{(-1)}$ had PRTF values above 1/e and the achieved resolution is 13.6 nm. A cross-section taken along the white line in Fig. 3(c) is shown in Fig. 3(d) to demonstrate that very sharp edges can be resolved by this setup. The distance between the points where the amplitude is 10% and 90%, commonly known as the 10/90 criterion, is 15 nm for the reconstruction object shown. The reconstructed phase (Fig. 3(e)) shows a flat profile with a linear gradient which is due to a shift of the center of the diffraction from the center of the camera and is commonly subtracted.

This result represents a record value for the resolution achieved with any table-top CDI setup. Note that the previous best resolution of 22 nm was found considering all spatial frequencies with PRTF value above zero [11]. Applying the established PRTF>1/e criterion [30] to our result and the previous record value, we find a factor of 3 improvement in resolution which was enabled by a unique combination of high photon flux and high coherence. The PRTF value is also above 0.2 at the corner of the camera (corresponding to 10 nm resolution) which indicates that sub-10 nm imaging is within reach with minor improvements of the HHG setup.

Another aspect of CDI, important for extending its applicability, is imaging at short acquisition times with meaningful resolution. This is useful in enlarging the field of view using techniques like ptychography where multiple partially overlapping diffraction patterns are recorded at slightly shifted sample positions to image larger sample areas [13,14], or 3D tomography, where multiple diffraction patterns are recorded at different angles [29]. Due to the very high photon flux, the presented imaging system allows high resolution CDI with integration times of only 3 sec. A corresponding diffraction pattern is shown in Fig. 5(a) with the object reconstructed from this measurement in Fig. 5(b).

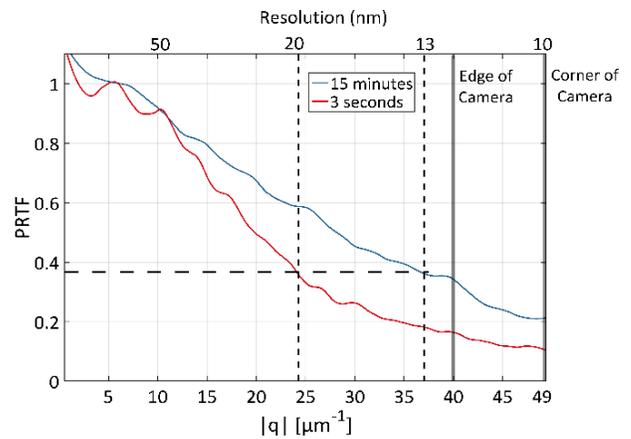

Fig. 4. PRTF and resolution plots of the high dynamic range (15 min.) and short acquisition time (3 sec.) measurements. The gray bars at |q|= 40 µm$^{-1}$ and 49 µm$^{-1}$ correspond to the edge and corner of the camera.

The diffraction pattern exhibits a good contrast until about half of the camera and the object is successfully reconstructed without loss of any features. The PRTF plot for this measurement is shown in Fig. 4 (red line) and the half-pitch resolution determined from the PRTF 1/e criterion was found to be 20.5 nm while the 10/90 resolution from the reconstructed object is 26 nm. This sub-30 nm resolution can now be achieved with only 3s of integration time, whereas previous table-top setups required many minutes – two order of magnitude longer – integration times [11,12].

In summary, we demonstrated a table-top CDI system providing a record high spatial resolution of 13 nm. This resolution is not only a factor of 3 better than previously reported, it is also the highest spatial resolution from any table-top XUV or X-ray imaging systems. The resolution is comparable to the latest high-end Zone-plate based X-ray microscopes operated at large scale facilities [1] which are e.g. employed for mask and chip inspection [3]. Our work demonstrates that a similar resolution is now available with a table-top setup. In addition, we demonstrated 20 nm resolution CDI with only 3 seconds of integration time. Thus, in future 3D tomography or ptychography on 100µm² sized samples, such as large biological cells, will be feasible further advancing table-top short-wavelength imaging. Due to the ever rising power of the driving lasers [31] and the harmonic sources [18], measurement times will be further reduced. In future shorter wavelengths [19] will push the spatial resolution to only a few-nanometers – far beyond the capabilities of zone-plate X-ray microscopy. Such "nanoscopes" on a table-top will complement electron- and visible light microscopy in researcher's laboratories and enable groundbreaking studies on the nanoscale in many areas of science as diverse as biology, optics, electronics, solid-state-physics and materials sciences.

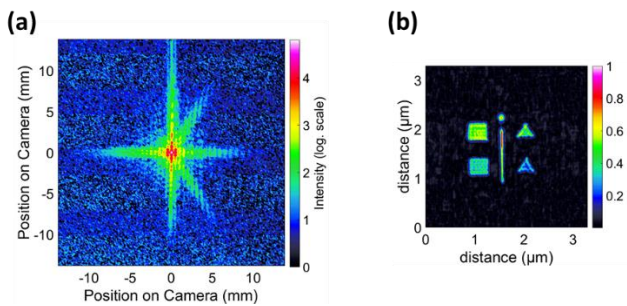

Fig. 5. (a) Diffraction pattern at 3 sec. of acquisition time with all other parameters kept constant. (b) Reconstructed object.

Finally, the femtosecond pulse durations of HHG light sources in principle allows for time-resolved observations of ultrafast-processes, such as nanoscale heat transport or ultrafast spin dynamics on their natural time scales. Such studies on smallest spatial and temporal scale will be of utmost importance for future high end electronics, storage devices, optics and materials.

**Funding**. We acknowledge support by the German Ministry of Education and Research (BMBF) - 05P2015 (APPA R&D: Licht-Materie Wechselwirkung mit hochgeladenen Ionen) and support by the Federal State of Thuringia and the European Social Fund - 2015 FGR 009: (XUV-Technologie und -Verfahren für Bildgebung mit nano-skaliger Auflösung (Nano-XUV)).

**Acknowledgement.** M. Zürch acknowledges support by the Alexander von Humboldt Foundation.